\documentstyle[12pt] {article}
\def\dspace{ \baselineskip = .30in}

\begin{document}

\title{Large $tan \beta$ from $SU(2)_R$ Gauge Symmetry}

\author{{\bf G. Lazarides and  C. Panagiotakopoulos}\\Physics
Division\\School of Technology\\University of
Thessaloniki\\Thessaloniki, Greece}

\date { }
\maketitle

\dspace
\vspace{30mm}
\centerline{\bf Abstract}
\vspace{10mm}

Sufficient conditions for the  relation
$tan \beta \simeq m_t/m_b$ to  hold in supersymmetric grand unified
theories  are formulated.
Essential ingredients  are the
$SU(2)_R$ gauge symmetry and a discrete matter parity.
The applicability of our conditions is illustrated by specific
examples.
Implications for neutrino masses are
discussed.

\newpage

The minimal supersymmetric standard model (MSSM), in spite of the
large number of undetermined parameters that it contains, is widely
believed to be the correct effective low energy theory of strong, weak
and electromagnetic interactions. One hopes, of course, that at least
some of the parameters  of this model will be
determined by embedding the standard
gauge group in a larger (unifying) one. This hope seems to be closer
to fulfillment as far as the well-known parameters $sin^2 \theta_w$
and $\alpha_s$ of the model are concerned. Assuming  a supersymmetry
breaking scale $M_s \sim 1$ TeV, the  renormalization group (RG)
equations of the MSSM are remarkably consistent$^{(1)}$ with the observed
values of  $sin^2 \theta_w$ and $\alpha_s$  and the unification
of the three gauge
couplings at a superheavy scale $\sim 10^{16}$ GeV.

The supersymmetric version of the standard electroweak theory,
however, although more successful than the non-supersymmetric one with
respect to "predicting" the ratio of the gauge couplings, introduces
an important undetermined new parameter by necessitating the doubling
of the number of electroweak higgs doublets. This new parameter, known
as $tan \beta$, is the ratio  of the
vacuum expectation values (VEVs) of the doublet $h^{(1)}$ giving mass
to the up-type quarks and the doublet $h^{(2)}$ giving mass to the
down-type quarks and charged leptons. The embedding of the MSSM in the
simplest supersymmetric grand unified theory (SUSY GUT), the minimal
SUSY SU(5) model, fails to determine $tan \beta$. However, SUSY GUTs based on
larger groups, like SO(10), may lead$^{(2)}$ to the asymptotic relation
$tan\beta = m_t/m_b$. In deriving this relation in SO(10), the
assumption is made
that the third generation fermions (mostly) acquire mass from a single
$16 \times 16 \times 10$ coupling with the electroweak higgs doublet
pair contained primarily in the 10-plet. This way the large mass of
the t-quark is explained without invoquing large ratios of yukawa
couplings within the third generation. Besides, detailed
investigations$^{(3)}$ have shown that large $tan \beta$ scenarios combined
with radiative electroweak breaking lead to very interesting
sparticle spectroscopy.

Our purpose here is to formulate, in the context of SUSY GUTs, widely
applicable conditions under which the  relation $tan \beta
\simeq m_t/m_b$ holds. A discrete matter parity (MP) and the $SU(2)_R$ gauge
group are essential ingredients of these conditions
which we apply to specific SUSY GUT examples.
In particular, we are
primarily
interested in embeddings of the MSSM into SUSY GUTs with semi-simple
gauge groups.

We consider SUSY GUT models which, below an energy scale $ M $, have
a symmetry group $P \times G$. $P$ is a global symmetry group and
$G \supseteq G_{LR} \equiv SU(3)_c \times SU(2)_L \times SU(2)_R
\times U(1)_{B-L}$ is a gauge group. At a scale $M_R \ll M$, $G_{LR}$ breaks
spontaneously down to the standard gauge group $G_{S} \equiv SU(3)_c
\times SU(2)_L \times U(1)_Y$. At energies  $ \sim M_s$, the unbroken
symmetry group contains $C \times G_{S}$, where $C$ is a  $Z_n$
symmetry. We will assume, without any essential loss of generality,
that the generator of $C$ is obtained by multiplying an element of $P$
with an element of the Cartan subgroup of $G$ commuting with $G_{S}$.
For the spectrum of the model, we assume that all
$G$-non-singlet states  are either heavier than $M$ or lighter than
$M^ \prime$ ($M>>M^ \prime \stackrel {_>} {_\sim} M_R$)
and the only states which are massless  in the  supersymmetric limit
are the states of the MSSM on  which $C$ acts as a  (generation blind)
$Z_n$ matter parity$^{(4)}$. The unique pair of electroweak higgs doublets
$h^{(1)}$,$h^{(2)}$ is assumed to form a single $SU(2)_R$ -doublet $h$.
The massive spectrum below $M$ should not contain any other states with
the $C \times G_{S}$ quantum numbers of the ordinary light quarks and
leptons. We will show that under the above conditions the approximate
relation $tan \beta \simeq m_t/m_b $ holds. If, in addition, (below
$M$)
there are
exactly  three $G_{S}$ -singlet states $\nu^c_i (i=1,2,3)$
which could possess Dirac mass terms with the ordinary
left-handed neutrinos,
then we also have $tan \beta \simeq
{m^D_{\nu_{\tau}}} /{m_ \tau}$ (with $m^D_{\nu_{\tau}}$  being the Dirac mass
of the $\tau$-neutrino).

\par
The states with masses $ \stackrel {_<}{_\sim} M^\prime$, on which we
focus, fall (up to corrections of order $M^\prime/M$ from heavier states)
into multiplets of  the gauge group $SU(2)_R$. Moreover, our
assumption about the generator of $C$ implies that, if one component
of a $SU(2)_R$ -multiplet happens to be an eigenstate of $C$, then all
its components are also eigenstates of $C$ (not necessarily with the
same eigenvalue).
Let $q_i (i=1,2,3)$
denote the  ordinary $SU(2)_L$ - doublet quarks and $u^c_i (d^c_i)$
the ordinary up (down) - type antiquarks, all of them being
massless superfields in
the supersymmetric limit. The $ q_i$ 's are assumed to be $SU(2)_R$
-singlets with $B-L=1/3$ and the $ u^c_i$ 's to belong
to $SU(2)_R$ -doublets with $B-L=-1/3$.
Let $d^{c \prime} _i$ be the partner of
$u^c_i$ in its
$SU(2)_R$ -doublet. The above discussion  shows that $d^{c
\prime}_i$, like $u^c_i$, is an eigenstate of $C$.
The existence of the terms $h^{(1)} q_i u^c_j (i,j = 1,2,3)$
implies (using the $SU(2)_R$ symmetry) the existence of the terms
$h^{(2)}q_i d^{c \prime}_j$ which combined with the assumed existence of the
terms $h^{(2)} q_i d^c_j$ leads to the conclusion that $d^c_i$ 's and
$d^{c \prime}_i$ 's have the same MP. Our assumption that  the only
states (below $ M$) with the $C \times G_{S}$ quantum numbers of
ordinary quarks and leptons are the  ordinary quarks and leptons
themselves
implies that (up to a simple renaming) the $d^{c \prime}_i$ 's coincide
with the $d^c_i$ 's. Therefore, the $u^c_i$ 's and the $d^c_i$ 's fall
by themselves into three $SU(2)_R$ -doublets, which we denote by $q^c_i$.
Analogous
arguments lead to the conclusion that the three ordinary light charged
antileptons  $e^c_i$ (assumed to belong to $SU(2)_R$ -doublets with
$B-L=1$)
and the three right-handed neutrinos $ \nu^c_i$
(provided there are exactly three such states) form three $SU(2)_R$
-doublets $ \ell^c_i$. Consequently, the tree-level light quark mass
terms come from couplings of the type $hq_i q^c_j$  while the tree-level
light charged lepton and neutrino Dirac mass terms from couplings of
the type $h \ell_i \ell^c_j$ ($\ell_i$ 's are the three ordinary light
lepton $SU(2)_L $-doublets). We see that the tree-level up and down quark
mass matrices are proportional with proportionality constant equal to
$tan \beta$. A similar proportionality holds between
neutrino Dirac mass matrices and charged
lepton mass matrices. Phenomenologically, this proportionality is
unacceptable for the quarks of the first two families. We assume that
for the first two families there should be considerable corrections to
the tree-level mass matrices from other sources whereas there should
be no such sizeable corrections for the third family$^{(5)}$. Thus, we
conclude that our scheme leads to the  relations mentioned
earlier.

In the special but very common case of a $Z_2$ MP, one
could relax the assumption that $h^{(1)}, h^{(2)}$ are partners in a
single $SU(2)_R$ -doublet $h$.It is enough to assume that
they just belong to $SU(2)_R$
-doublets and that  (below $M$) there are no other states  which belong
to $ SU(2)_R$ -doublets and carry
the same $C \times G_{S}$ quantum numbers as $h^{(1)}, h^{(2)}$.
We remind the reader that under the $Z_2$
MP the electroweak higgs doublet pair remains invariant
while the ordinary quarks and leptons change sign. Let $h^{(2)
\prime}$ be the partner of $h^{(1)} $ in its $SU(2)_R$ -doublet. It is
easily seen that, under $G_{S}, h^{(2)\prime }$ and $\ell_i$ 's  have the same
quantum numbers. Then, if $h^{(2) \prime}$ has different MP from
$h^{(2)}$, namely minus, it must coincide with one of the light
leptons $ \ell_i$. This is, of course, impossible because the $ \ell_i$
's are not $SU(2)_R$ -doublets. Therefore, $h^{(2) \prime}$
belongs to a $SU(2)_R$ -doublet and is identical under $C \times
G_{S}$ with $h^{(2)}$. Consequently, $h^{(2) \prime}$
coincides with $h^{(2)}$. Thus, $h^{(1)}$ and $h^{(2)}$ are partners
in a single $SU(2)_R$ -doublet $h$.

Finally, even in the general case of a $Z_n$ MP, there are
situations in which
one could relax the assumption that $h^{(1)}, h^{(2)}$ belong to
the same
$SU(2)_R$ -doublet.
In fact, no assumption whatsoever on the $SU(2)_R$ properties of
$h^{(1)}, h^{(2)}$ is needed
provided (below $M$) there are no other states
identical under $C \times G_{S}$ with them
and the light $u^c_i$'s, $d^c_i$'s form $SU(2)_R$ -doublets $q^c_i$.
The last requirement is readily fulfilled if there are just three
states with the $G_{S}$ quantum numbers of a $d^c$.
The
existence of the terms $h^{(1)} q_i u^c_j$ together  with the
uniqueness of $h^{(1)}$ under $C \times G_{S}$
imply that $h^{(1)}$ belongs to a $SU(2)_R$ -doublet $h$.
Let $ h^{(2) \prime}$
be the partner of $ h^{(1)}$ in $h$.
Then the terms $h^{(2) \prime} q_i d^c_j$ are certainly allowed  (by
$SU(2)_R$ symmetry) leading to the conclusion that $h^{(2) \prime}$ and
$h^{(2)}$ are indistinguishable under $C \times G_{S}$ and,
consequently, identical. Thus, $h^{(1)}$ and $h^{(2)}$ are partners in
a single $SU(2)_R$ -doublet $h$.

We should point out here that global symmetries  acting as
matter parities are standard ingredients of supersymmetric models.
They are introduced  in order to deal with the  well-known problem of
rapid proton decay through $d=4$ operators at the level of the
supersymmetric standard model. Therefore, our assumption that the
gauge group $G_{S}$ is supplemented with a MP  $C$ should not
at all be considered as an extra restriction.

As a first application, we consider  a SUSY GUT model  with a $P \times
G$ symmetry, where $P$ is a $Z_2$ global symmetry  and the gauge group $G
\equiv$ SO(10) $\supset  G_{LR}$.  The scale $M$ here could be
taken to be the  Planck mass $M_P \sim 10^{19} $ GeV. $G$ breaks  down
at a scale $M^ \prime \equiv M_R \sim 10^{16}$ GeV directly to
$G_{S}$ using higgs fields
with  $P=+1$  in the $16, \overline {16}, 126, \overline {126}$, 45,
54 and 210 representations of  SO(10). The
three ordinary light generations are contained in three 16-plets with
$P = -1$.
There are also  10,
$ \overline {126} $ and 120 representations
of SO(10) with
$P=+1$. We assume that the electroweak higgs doublets
$h^{(1)}, h^{(2)}$ are partners in a single $SU(2)_R$ -doublet
and belong to an otherwise arbitrary linear combination of the 10,126,
$\overline {126}$ and 120 representations. Higgs fields in the 16,
$\overline{16}$ and 210 representations could also contribute to $h^{(1)},
h^{(2)}$. These contributions, however, cannot belong to a single
$SU(2)_R$ -doublet and, thus, should be suppressed.
Notice that  $P$ remains unbroken by
all the VEVs and coincides with the  MP $C$.
Furthermore, there are no other states having the same $C \times
G_{S}$ quantum numbers with  the light quarks and
leptons and there are just three $\nu^c_i$'s.
Thus, we conclude that all the conditions for  the relation
$tan \beta \simeq  {m_t} / {m_b} \simeq
{m^D_{\nu_{\tau}}} / {m_ {\tau}}$ to hold are satisfied. It is
important to note that the validity of this relation is not automatic
even in the minimal SUSY SO(10) model which contains higgs fields in
the 16, $\overline{16}$ representations.

To construct an example with automatically large $tan \beta$ and the
usual MSSM predictions of $sin^2 \theta_w$ and $\alpha_s$, we consider
a model with symmetry $P_1 \times P_2 \times P_3 \times G$, where
$P_1, P_2$ and $P_3$ are $Z_2, Z_2$ and $Z_7$ global symmetries
respectively and $G$ the  gauge group $SU(4)_c \times SU(2)_L \times
SU(2)_R$. The scale $M$ is a superheavy scale close to $M_P$, where we
assume a common value for the gauge couplings of the three factors of
$G$. Let $A_L, A_R, D, B, H$ and $T$ denote left-handed fields
transforming under $G$ as (4,2,1), ($\bar {4}$,1,2), (6,1,1), (15,1,1), (1,2,2)
and (1,1,1) respectively.
The field content of the model  consists of
three $A_L (-1, +1, \alpha)$, three $A_R (-1, +1, \alpha)$, three
$D(+1,+1,1)$, one $B(+1,+1,1)$, one $H(+1,+1, \alpha^5)$, one $T(+1,+1,
\alpha^3)$, one $A_L(+1,-1, \alpha)$, one $\bar {A}_L (+1, -1,
\alpha^6)$, one $A_R(+1, +1, \alpha)$ and one $\bar{A}_R(+1,+1,
\alpha^6)$, where the transformations of the various fields under the
generators of $P_1,P_2, P_3$ are given in parenthesis and $\alpha =
e^{2 \pi i/7}$.
The large VEVs, all of the order of $M_R \simeq 10^{16}$
GeV, are acquired by the fields $B(+1, +1, 1)$, $A_R(+1,+1, \alpha)$,
$\bar{A}_R(+1,+1,\alpha^6)$ and $T(+1,+1, \alpha^3)$. As a result, the
gauge group $G$ breaks down to $G_{S}$ and the symmetry $P_3$ breaks
down completely. The symmetry $P_1 \times P_2$ remains unbroken and is
identified with the MP  $C$. All the allowed tree-level mass
terms are assumed $\sim M_R \equiv M^ \prime$. Below the scale $M_R$, we
recover the
MSSM supplemented with exactly three right-handed neutrinos (mass
$\sim M^2_R/M$). Ordinary light quarks and leptons are contained in
the $A_L$ 's and $A_R$ 's with $C$-charge ($ -1, +1$) and are unique
under $ C \times G_{S}$. Besides, $h^{(1)}$ and $h^{(2)}$ are the
$SU(2)_R$ partners  in $H$. Their mass is $\sim <T>^6 /M^5
\sim M_s$. Therefore, $tan \beta$ is automatically large. The role of
the symmetry $P_3$ is  to protect $H$ from acquiring a mass
$\sim M_R$. It also suppresses the rate of proton decay mediated by the
$D$'s. Due  to the appropriately chosen spectrum, the one-loop RG
equations above $M_R$ predict identical running for the gauge
couplings of the  groups $SU(4)_c, SU(2)_L $ and $SU(2)_R$.
Combining this fact with their assumed equality at $M$, we conclude
that equality of the gauge couplings of $G_{S}$ at the scale $M_R$ is
a justified boundary condition. The successful MSSM predictions for
$sin^2 \theta_w$ and $\alpha_s$ follow immediately.

As another example, we consider a SUSY model with  symmetry group $P_1
\times P_2 \times G$. $P_1$ is a $Z_2$ global symmetry which remains
unbroken and coincides with the $Z_2$ MP $C$, $P_2$ is a
$Z_6$ global symmetry
and $G$ coincides
with the gauge group $G_{LR}$.
The coupling constants  of the four factors in
$G_{LR}$ are assumed to become equal  at a scale $M_c = 2.4 \times
10^{18}$ GeV  probably related to a more
fundamental theory.
The spectrum below $M_c$ contains the three quark, $q_i$, antiquark,
$q^c_i$, lepton, $\ell_i$, and antilepton, $\ell^c_i$, fields, one pair
of colorless
$SU(2)_L$ -doublet fields, $H^{(1)}, H^{(2)}$, which form a $SU(2)_R$
-doublet
with zero $ (B-L)$-charge,one $\ell_o, \bar{\ell}_o$ pair,
one $\ell^c_o, \bar{\ell}^c_o$ pair,
eight $q_m , \bar{q}_m$ and $q^c_m , \bar{q}^c_m$ ($m$ =1,...,8)
pairs and one singlet $T$.
All fields, except the  ordinary quark and lepton generations, have
$P_1=+1$. Under the $P_2$ symmetry generator all fields get
multiplied by
$ \alpha = e^{2 \pi i/6}$  except $H^{(1)}, H^{(2)}$ which get
multiplied by
$ \alpha^4$ and $ \bar {\ell}_o, \bar{\ell}^c_o, \bar {q}_m, \bar {q}^c_m$
by $ \alpha^5$. The $\ell^c_o , \bar {\ell}^c_o$ pair is
assumed to have appropriate superpotential couplings in order to break
the $SU(2)_R \times U(1)_{B-L}$ gauge symmetry down to $U(1)_Y$ at a
scale $M_R = 10^{15}$ GeV.
All other allowed superpotential mass terms are
assumed to be equal to $M = 2.4 \times 10^{17}$ GeV.

In the absence of $P_2$, there are  no light states
which could play the role of the electroweak higgs doublets.
This is due to the fact that
the only states able to play such a role, namely $H^{(1)},
H^{(2)}, \ell_o$ and $ \bar{\ell}_o$, acquire tree-level masses
$\sim M$. $P_2$, however, allows the mass terms $<\nu^c>
H^{(1)} \ell_o$ and  $M \bar {\ell}_o \ell_o$ but forbids the mass
terms $M H^{(1)} H^{(2)} $ and $<\bar {\nu}^c > \bar {\ell}_o
H^{(2)}$. As a result, there is an electroweak higgs doublet pair
$h^{(1)} \simeq \rho [ H^{(1)} - ({M_R}/{M}) \bar {\ell}_o],
h^{(2)} =H^{(2)}$ (where $\rho = [1 + ({M_R}/{M})^2]^{-1/2})$ left
massless with the orthogonal states $\rho [ \bar{\ell}_o + (
{M_R}/{M}) H^{(1)}]$ and $\ell_o$ acquiring superheavy masses $\sim M$.
$P_2$
must, of course, be broken in order for a higgsino mass term $\sim M_s$
to be generated.
A VEV  $<T> \simeq \frac{M_R}{2}$ achieves this breaking. The
superpotential coupling $H^{(1)} H^{(2)} <T^4>/ M^3_c$  generates,
then, a mass term with the desired order of magnitude without
significantly altering the above discussion.

It is easily seen that the only states below the scale $M \gg M_R
\equiv M^ \prime$ are
the states of the MSSM supplemented with three
right-handed neutrinos (with
masses  $\sim M^2_R/M_c$), the massive gauge  supermultiplet associated
with the generators of $SU(2)_R \times U(1)_{B-L} /U(1)_Y$ and the
higgs associated with the breaking of $SU(2)_R \times U(1)_{B-L}$ down
to $U(1)_Y$ with mass  $\sim M_R$. Notice that (below $M$) the states
of the MSSM are unique under $C \times G_{S}$ and, moreover, there
are only three states with the $G_{S}$ quantum numbers of a $d^c$.
Our relation for $tan \beta$ follows
immediately.

We can easily estimate the corrections to the
relation for $tan \beta$ due to the spectrum above $M$. It should be
clear that the only sizable correction could come from the fact that, due to
the small $ (\sim M_R/M)$ contribution of $ \bar{\ell}_o$ in
$h^{(1)},h^{(1)}$
is not exactly the $SU(2)_R$ -partner of $h^{(2)}$.
The exact
relation is ${m_t}/{m_b} = <H^{(1)}> /<H^{(2)}>$ instead of
${m_t}/{m_b} = <h^{(1)}>/<h^{(2)}>$ because $H^{(1)}$, and not $\bar
{\ell}_o$, couples to quarks and leptons. Using the readily obtainable
relation $<H^{(1)}> = \rho <h^{(1)}>$, we get $m_t/m_b =
\rho$ $tan \beta$. It seems that in this case the departure of the
coefficient $\rho$ from unity is $\simeq \frac{1}{2} (
{M_R}/{M})^2$ and not $\sim M_R/M$ as one would naively have thought.

For the sake of completeness we mention that our simple model with the
chosen field content is consistent with the presently favored values
for $sin^2 \theta_w$ and $ \alpha_s$. An  one-loop calculation gives the
perfectly acceptable values  $sin^2 \theta_w =0.234$ and $ \alpha_s
(M_Z)
= 0.120$ with the perturbative value $ \alpha_G=0.06$ for the unified gauge
structure constant at $M_c$.

Our last  example concerns a natural and very
attractive model based on the gauge group $\tilde {G} \equiv
SU(3)_c \times SU(3)_L
\times SU(3)_R$. An advantage of SUSY $SU(3)^3 $ models
over the usual GUTs is
that
the proton decay amplitude through $d=5$ operators is not
related to the light quark masses and, thus, can
be suppressed
through suitable discrete symmetries.
The present model, a variation of the model in ref.(6),
possesses a
symmetry group $P \times \tilde{G}$ with $P \equiv P_1 \times P_2
\times P_3$   a $Z_2
\times Z_2 \times Z_2$ global symmetry.
We assume a common
unified  gauge coupling at a scale $M_c=2.4 \times 10^{18}$ GeV as
before. The model makes use only of gauge singlets, $\lambda =
(1, \bar{3}, 3), Q = (3,3,1)$ and $Q^c (\bar{3},1,\bar{3})$ fields and
their mirrors.
The only
difference from the model of ref.(6) is that we omit one $P$ -invariant
$\lambda , \bar{\lambda}$ pair
in order to obtain the correct values of $sin^2
\theta_w$ and $ \alpha_s$ since we now assume a two step symmetry
breaking chain. Therefore, there are seven $\lambda$ 's, eight $Q$ 's
and $Q^c$ 's, four $\bar {\lambda}$ 's and five $\bar {Q}$ 's and
$\bar{Q}^c$ 's.  Under $P_1$, all color singlets remain
invariant while all color triplets and antitriplets change sign. Under
$P_2$, all fields remain invariant except  $\lambda_4, \lambda_7, \bar
{\lambda}_2$ and $\bar {\lambda}_4$ which change sign. Finally, under
$P_3$, all fields remain invariant except $\lambda_4, \lambda_5,
\lambda_6, \bar{\lambda}_2, \bar{\lambda}_3, Q_1, Q_2$ and $Q^c_3$
which change sign.
At the scale $M_X$, the symmetry $\tilde{G}$ breaks down to
$G \equiv G_{LR}$(by the VEVs $\mid < \lambda_6 > \mid = \mid <\bar
{\lambda}_3 > \mid$)
which, in turn, breaks down to $G_{S}$ (by the VEVs $\mid < \lambda_7
>\mid = \mid < \bar {\lambda}_4 > \mid$)at
a scale $M_R$. Moreover, a $Z_2$ MP  $C$ (the combination of
$P_2$ with the center of $SU(2)_R$)
commuting with
$G_{S}$ survives as an exact  symmetry. All
$\tilde{G}$ -non-singlet states acquire masses  $\sim M_X$ except  the
states of the MSSM together with three right-handed neutrinos (with mass
$\sim M^2_R/M_c)$, the  massive gauge supermultiplet associated with
the generators of $SU(2)_R \times U(1)_{B-L} /U(1)_Y$ together with
the higgs fields necessary for the breaking of $SU(2)_R \times
U(1)_{B-L}$ down to $U(1)_Y$ and  one $\ell , \bar{\ell}$ pair with a
mass  $\sim M^2_X/M_c$. This pair is identical under $C \times G_{S}$
with the
$h^{(1)}, h^{(2)}$ pair. The situation coincides with the one
encountered in the previous example after identification of $M$ with
$M^2_X/M_c$. The only question that remains to be answered is whether
there are values of $M_X$ and $M_R$ consistent with the measured
values of $sin^2 \theta_w$ and $ \alpha_s$ and satisfying the inequality
$M^\prime \equiv M_R<< M \equiv M^2_X/M_c$.
It turns out that, for  $M_X=10^{17.1}$ GeV and
$M_R=10^{15}$ GeV, the one-loop RG equations give
the perfectly acceptable values $sin^2 \theta_w
(M_Z) = 0.232$ and  $ \alpha_s (M_Z) = 0.121$. For the above values,
the expected departure from unity of the coefficient $ \rho$, in the
relation $m_t/m_b = \rho \> tan \beta$, is only $\simeq
\frac{1}{2} (M_R/M)^2 =0.0126$ which is sufficiently small.

It is, perhaps, appropriate at this point to make some remarks
concerning the implications of the relation $tan \beta \simeq m_t/m_b
\simeq  {m^D_{\nu_\tau}}/{m_{\tau}}$ for neutrino masses. The large
value of $tan \beta$ implied by this relation forces
$ m^D_{\nu_{\tau}}$ $ \sim 100$ GeV. Consequently, in order for
$m_{\nu_{\tau}}$ to satisfy the cosmological
bounds, the Majorana mass of
$\nu^c_\tau$ $ \stackrel{_>}{_\sim} 10^{12}$ GeV. This fact certainly places
a lower bound on the $SU(2)_R$ symmetry breaking scale $M_R$. In the
very common case in which the Majorana $\nu^c$ mass is due to
non-renormalizable  terms,
this bound on $M_R $ is $ \sim 10^{15}$ GeV. $\nu^c$ Majorana
masses typically $\sim 10^{12}$GeV and  $m^D_{\nu_{\tau}}$ $\sim 100$
GeV  offer the exciting possibility that
$\nu_\tau$'s
contribute significantly to the "hot" component of the dark
matter of the universe. At the same time the two lighter neutrinos
could have masses small enough  to be consistent with the MSW
solution of the solar neutrino problem.

In summary, we investigated the possibility that the well-known free
parameter $tan \beta$ of the MSSM can be determined by embedding MSSM in a
SUSY GUT. Sufficient conditions for the relation $tan \beta \simeq
m_t/m_b \simeq {m^D_{\nu_{\tau}}}/{m_{\tau}}$ to hold were given
and the importance of the $SU(2)_R$ gauge symmetry in
this connection was emphasized. We discussed examples of such SUSY
GUTs in some detail with particular emphasis on semi-simple gauge
groups.
\newpage

\section*{References}

\begin{enumerate}
\item J. Ellis, S. Kelley and D.V. Nanopoulos, Phys. Lett.
\underline{B249} (1990) 441;
V. Amaldi, W. de Boer and H. Furstenan, Phys. Lett. \underline{B260}
(1991) 447;
P. Langacker and M.X. Luo, Phys. Rev. \underline{D44} (1991) 817.

\item B. Ananthanarayan, G. Lazarides and Q. Shafi, Phys. Rev.
\underline{D44} (1991) 1613;
H. Arason, D.J. Casta\~{n}o, B.E. Keszthelyi, S. Mikaelian, E. J. Piard,
P. Ramond and B.D. Wright, Phys. Rev.Lett. \underline{67} (1991) 2933;
S.Kelley, J.L.Lopez and D.V. Nanopoulos, Phys. Lett
\underline{B274}(1992) 387.

\item B. Ananthanarayan, G. Lazarides and Q. Shafi, Phys. Lett.
\underline{B300} (1993)245;
V.Barger, M.S. Berger and P. Ohmann, Wisconsin Univ. preprint
MAD/PH/801 (1993).

\item L.E. Ib\'{a}\~{n}ez and G.G.Ross, Nucl.Phys.
\underline{B368} (1992)3.

\item L.J.Hall, R. Rattazzi and U.Sarid, LBL preprint 33997(1993).

\item G. Lazarides and C. Panagiotakopoulos, Thessaloniki Univ.
preprint UT-STPD-2-93(1993).
\end{enumerate}

\end{document}